\documentclass{article}
\usepackage{amsmath,amsthm}
\usepackage{amssymb,latexsym}
\usepackage[mathscr]{eucal}
\usepackage{setspace}
\usepackage{graphics}
\usepackage{array}

\newcommand{\pa}{\partial}
\newcommand{\p}{^\prime}

\newcommand{\al}{\alpha}

\newcommand{\dpr}{^{\prime\prime}}

\newcommand{\pr}{^\prime}

\newcommand{\rp}{\right)}
\newcommand{\lp}{\left(}
\newcommand{\rb}{\right]}
\newcommand{\lb}{\left[}
\newcommand{\rel}{\right\}}
\newcommand{\lel}{\left\{}

\newcommand{\tothehalf}{^\frac{1}{2}}
\newcommand{\beq}{\begin{equation}}
\newcommand{\eq}{\end{equation}}

\newcommand{\inta}{\int_A}

\newcommand{\ovq}{\overline{q}}

\newcommand{\ovp}{\overline{\phi}}
\newcommand{\hatq}{\hat{q}}
\newcommand{\hatp}{\hat{\phi}}
\newcommand{\C}{\xi}

\usepackage{graphicx}
\graphicspath{{converted_graphics/}}
\begin{document}

\title{\textbf{Zonal Shear Flows with a Free Surface: Hamiltonian Formulation and Linear and Nonlinear Stability}}         
\author{B.K. Shivamoggi\footnote{\large {Permanent Address: University of Central Florida, Orlando, FL 32816-1364}} and G.J.F. van Heijst\\
J. M. Burgers Centre and Fluid Dynamics Laboratory\\
Technische Universiteit Eindhoven\\
NL-5600 MB Eindhoven\\
The Netherlands}        
\date{}          
\maketitle
\large{\bf Abstract}

General theoretical results via a Hamiltonian formulation are developed for zonal shear flows with the inclusion of the vortex stretching effect of the deformed free surface. These results include a generalization of the Flierl-Stern-Whitehead {\it zero angular momentum theorem} for localized nonlinear structures (on or off the $\beta$-plane), and {\it sufficient} conditions for {\it linear} and {\it nonlinear} stability in the Liapunov sense - the latter are derived via bounds on the equilibrium potential vorticity gradient.

\pagebreak

\section{Introduction}       
Numerical simulations (Marcus \cite{Mar}), laboratory experiments (Sommeria et al. \cite{Som}, Chomaz et al. \cite{Cho}) and analytical treatments (Shivamoggi \cite{Shi1}) indicated that, in a quasi-geostrophic flow, the existence of a strong zonal-flow shear can lead to the formation of a strong persistent coherent vortex. On the other hand, large scale wave motions, or meanders, occurring in the oceanic circulations are believed to tbe caused by the instability of the zonal shear flow (Kuo \cite{Kuo1}). On this premise, the linear stability of a barotropic zonal shear flow render the influence of differential rotation (as represented by the Rossby parameter $\beta$) was addressed by Kuo \cite{ Kuo2}, Lipps \cite{Lip}, Pedlosky \cite{Ped1} and \cite{Ped2}, Howard and Drazin \cite{How}, Drazin et al. \cite {Dra}, Hickernell \cite{Hic}, Gnevyshev and Shrira \cite{Gne}, and others. These investigations were generalized (Ripa \cite{Rip}, Shivamoggi and Rollins \cite{Shi2}) to include the vortex-stretching effect of the deformed free surface. In this paper, we give a Hamiltonian formulation for this problem, and use this framework to give -

\begin{enumerate}
  \item a generalization of the Flierl-Stern-Whitehead \cite{Fli} {\it zero angular momentum theorem} for localized nonlinear structures (on or off the $\beta$-plane);
  \item a {\it sufficient} condition for {\it linear} stability in the Liapunov sense for steady states of this system that complements the previous result of Ripa \cite{Rip};
  \item {\it sufficient} conditions for {\it nonlinear} stability in the Liapunov sense via bounds on the equilibrium potential vorticity gradient.
\end{enumerate}

\section{Hamiltonian Formulation}
Consider a two-dimensional {\it quasi-geostrophic} flow in which the baroclinic effects are produced by the deformed free surface of the ocean. The governing equation (in appropriate units) is (Charney \cite{Cha}

\beq
\frac{\pa q}{\pa t} + [\phi, q]=0
\eq

\noindent
where $\phi$ is the stream function and $q$ is the potential vorticity - 

\beq
q \equiv \nabla^2\phi-\phi+f
\eq

\noindent
f being the Coriolis parameter and,

\beq
[A,B]\equiv \frac{\pa A}{\pa x}\frac{\pa B}{\pa y} - \frac{\pa A}{\pa y}\frac{\pa B}{\pa x}.
\eq

\noindent
The Hamiltonian for this system is 

\beq
H=-\frac{1}{2}\int_A \phi q dA.
\eq

\noindent
assuming that $\phi=0$ on the boundary C enclosing the area $A$ where the flow occurs.

Take $q$ to be the canonical variable and the skew-symmetric operator $J$ to be 

\beq
J\equiv -[q,(\cdot)].
\eq

The Hamilton equation is then given by 

\beq
\frac{\pa q}{\pa t} = J\frac{\delta H}{\delta q} = [q,\phi]
\eq

\noindent
which is just equation (1), as required!

The Casimir invariants for this system are the solutions of 

\beq
J\frac{\delta C}{\delta q} = -\lb q,\frac{\delta C}{\delta q}\rb=0
\eq

\noindent
which implies

\beq
\frac{\delta C}{\delta q} = F\p  (q)
\eq

\noindent
or

\beq
C=\int_A F(q) dA.
\eq

$C$ may be understood by noting that $q$ is actually a Lagrange invariant, so any integral function of $q$ would be conserved!

\section{The Beltrami States}
The Beltrami state is the minimizer of H keeping $C$ constant, and is given by 

\beq
\frac{\delta H}{\delta q} = \lambda \frac{\delta C}{\delta q}
\eq

\noindent
or

\beq
-\phi = \lambda F\p (q)
\eq

\noindent
or

\beq
q=P(\phi).
\eq

\section{Translational Symmetry Solutions}
If this system has translational symmetry along the x-direction, the latter is characterized by an invariant $M$ given by 

\beq
J\frac{\delta M}{\delta q} = -\frac{\pa q}{\pa x}
\eq

\noindent
which implies

\beq 
M= \int_A yq dA.
\eq

On the other hand, existence of a traveling wave solution for this system implies that

\beq
q (x,y,t) = q(\C,y), \C\equiv x-ut
\eq

\noindent
which means, in turn,

\beq
-u\frac{\pa q}{\pa x}=\frac{\pa q}{\pa t}.
\eq

On using equations (6) and (13), equation (16) implies 

\beq
uJ\frac{\delta M}{\delta q}=J\frac{\delta H}{\delta q}
\eq

\noindent
or 

\beq
J\frac{\delta}{\delta q} (H-uM)=-\lb q,\frac{\delta}{\delta q} (H-uM)\rb=0.
\eq

On using (4) and (14), equation (18) leads to 

\beq
[q, \phi-uy]=0.
\eq

\section{Generalization of Flierl-Stern-Whitehead Zero Angular Momentum Theorem}

Multiply equation (19) through by $\C$, and integrate over A; this leads to 

\beq
\int_A [q, \phi-uy]\C dA = 0.
\eq

Assuming localized traveling structures, (20) reduces to

\beq
\int_A (\phi-uy)q_y dA=0
\eq

\noindent
or

\beq
\int_A \phi f\p dA +u\inta q dA=0.
\eq

Assume Beltrami states given by

\beq
q=a\phi
\eq

\noindent
$a$ being a constant, which is a special case of (12) and is of added interest because it then yields an $H$ which is quadratic in $\phi$! The Beltrami condition is equivalent to the {\it ``slow variation"} restriction used by Flierl, Stern, and Whitehead (FSW) \cite{Fli} and quantifies it precisely. 

On using (23), equation (22) becomes

\beq
\inta (au+f\p)\phi dA=0
\eq

\noindent
which is a generalization of the FSW {\it zero angular momentum theorem} \footnote{\large The net angular momentum is 
\beq\large\notag
\begin{array}{lll}\Large
L &=&\iint_A(xv-yu)dxdy\\
  &=-&\iint_A(x\phi_x + y\phi_y)dxdy\\
  &=&\iint_A 2\phi dxdy.
\end{array}\eq
}
\noindent
for localized nonlinear structures that results in the $\beta$-plane approximation $(f\p=const\equiv\beta)$-

\beq
\inta \phi dA=0.
\eq

Note that in the generic case, more general localized nonlinear structures with {\it non-zero} angular momentum are possible, especially off the $\beta$-plane! On the other hand, even on the $\beta$-plane, observe that the traveling localized nonlinear structures are not restricted by the FSW theorem if they satisfy the critical condition - 

\beq
u=-\frac{\beta}{a}.
\eq

This condition, it is to be noted, is not restrictive because the constant $a$ is arbitrary. It may be noted that the present formulation could be trivially generalized to include zonal shear flows.

\section{Linear Stability in the Liapunov Sense}

The stationary state of the system in question is a stationary point of the functional - 

\beq
I[q]\equiv H[q]+\C[q] = \inta \lb-\frac{1}{2} \phi q + F(q)\rb dA
\eq

\noindent
and is given by the vanishing of the first variation of $I[q]$ - 

\beq
\delta I[q]=\inta[-\overline{\phi}+F\p(\overline{q})]\delta q\  dA = 0
\eq

\noindent
from which,

\beq
\overline \phi = F\p(\overline{q})\  or\  \overline{q}=P(\overline{\phi}).
\eq

Next, the second variation of $I[q]$ is given by

\beq
\delta^2 I[q]=\inta\lb\frac{1}{2}(\nabla\delta\phi)^2 + \frac{1}{2}(\delta\phi)^2+\frac{1}{2}F\dpr(\overline{q})(\delta q)^2]\rb dA.
\eq

A sufficient condition for {\it linear} stability in the Liapunov sense (Arnol'd \cite{Arn}) is that $\delta^2 I$ is positive (or negative) definite for all suitably smooth variations $\delta q.$ This requires, from (29), that 

\beq
F\dpr (\overline{q}) \geq 0 , \ \ \forall (x,y) \in A
\eq

\noindent
which, on using equation (28), implies

\beq
\frac{d\overline{\phi}}{d \overline{q}}\geq0
\eq

\noindent
or

\beq
\lp\frac{\pa \overline{q}}{\pa y}\rp\lp\frac{\pa \overline{\phi}}{\pa y}\rp \geq 0, \ \ \ \forall\ (x,y)\ \in\ A.
\eq

If the system has translational symmetry along the x-direction, then (32) becomes 

\beq
\lp\frac{\pa \overline{q}}{\pa y}\rp\lp\frac{\pa}{\pa y}(\overline{\phi}+cy)\rp \geq0
\eq 

For the parallel zonal shear flows, $\overline{u}=\overline{u}(y),$ the {\it linear} stability criterion (33) becomes 

\beq
(\overline{u}\ \dpr - \beta-\overline{u})(\overline{u}-c)\geq0
\eq

\noindent
with the equality (where the potential vorticity attains its maximum) valid at $y=y_c$ where $u\!\!=\!\!c$ - in agreement with the {\it normal-mode} result (Shivamoggi and Rollins \cite{Shi2}). In the absence of the planetary rotation, this result reduces to Fjortoft's theorem. 

The above result was also given by Ripa \cite{Rip} by using an energy method that made use of invariants associated with explicit symmetries of the system. It may be noted that the present development complements this formulation by using instead Casimir invariants associated with the non-canonical nature of the Eulerian formulation of the system. 

Note that the stability criterion (34) is applicable to an arbitrary perturbation $q$ which allows $I[q]$ to exist. Thus the normal-mode result is generalizable to arbitrary perturbations and implies even the normed stability:

{\bf THEOREM}: The steady state $(\ovp, \ovq)$ is linearly stable in the Liapunov sense with respect to the perturbation norm - 

\beq
||(\delta\phi, \delta q)|| \equiv \lel \delta^2 I [\ovq]\rel\tothehalf
\eq

\noindent
if the Casimir function $F(q)$ satisfies 

\beq\tag{31}
F\dpr (q)\geq 0,\ \ \ \forall\ (x,y)\ \in\ A.
\eq

\section{Nonlinear Stability}

{\it Nonlinear} stability theory seeks to establish stability of a steady state by shown that it is a local extremum of energy with respect to {\it finite}-amplitude, Casimir-preserving perturbations. In order to address nonlinear stability, one needs to examine the sign of the {\it exact} invariant (Arnol'd \cite{Arn}) - 

\beq
\Delta I[q]\equiv I[q]-I[\ovq]
\eq

\noindent
instead of $\frac{1}{2} \delta^2 I$ which was considered in the {\it linear} stability problem. If $\Delta I$ is positive (or negative) definite, then it can be used to define a perturbation norm, and the steady state (29) is {\it nonlinearly} stable in this norm. This development needs additional {\it convexity} conditions to be imposed on the functional $I[q]$ in order to insure that the latter is {\it convex} in a small but {\it finite} neighborhood of the steady state (29).

Let us write $\Delta I[q]$ as follows - 

\beq
\Delta I[q]=I[\ovq+\hat{q}]-I[\ovq]
\eq

\noindent
where $\hat{q}$ is the {\it finite}-amplitude perturbation imposed on the stationary state $\ovq$.

Using (27) and (29), we have,

\beq
\Delta I [\hat{q}]=-\inta\frac{1}{2}\hat{\phi}\hatq dA+\inta[F(\ovq+\hatq)-F(\ovq)-F\pr(\ovq)\hatq]dA.
\eq

Note,

\beq
\Delta I[\hatq]=\delta^2I + 0(\hatp^3,\hatq^3,\hatp^2\hatq,\hatp\hatq^2)
\eq

\noindent
so, $\delta^2 I$ is only the {\it small}-amplitude approximation to $\Delta I$!

Suppose $F(q)$ satisfies the {\it convexity} condition - 

\beq
0<\al_1<F\dpr(q)<\beta_1<\infty, \ \forall q
\eq

\noindent
where $\al_1$ and $\beta_1$ are real numbers. So, 

\beq
\frac{1}{2}\al_1\hatq^2<[F(\ovq+\hatq)-F(\hatq)-F\pr(\ovq)\hatq]<\frac{1}{2}\beta_1\hatq^2
\eq

\noindent
and hence, from (39), we have

\beq
\frac{1}{2}\inta[-\hatp\hatq+\al_1\hatq^2]dA<\Delta I[\hatq]<\frac{1}{2}\inta[-\hatp\hatq+\beta_1\hatq^2]dA.
\eq

\noindent
which shows that the {\it convexity} condition (41) is {\it sufficient} to establish the positive (or negative) definiteness of the {\it exact} invariant $\Delta I[q].$ Further, (43) implies that a perturbation norm defined by 

\beq
||\hatq||\equiv \{ \Delta I^*[\hatq]\}\tothehalf
\eq

\noindent
where, 

\beq
\Delta I^*[\hatq]\equiv\frac{1}{2}\inta[-\hatp\hatq+\al_1\hatq^2 ]dA
\eq

\noindent
is bounded from above.

{\bf THEOREM}: Suppose that the Casimir function $F(q)$ satisfies the following {\it convexity} condition - 

\beq\tag{41}
0<\al_1<F\dpr(q)<\beta_1<\infty,\ \forall q
\eq

\noindent
for some real constants $\al_1$ and $\beta_1$. Then, the stationary state $\ovq$, determined by 

\beq\tag{29}
\overline{\phi}=F\pr(\ovq)
\eq

\noindent
is {\it nonlinearly} stable in the Liapunov sense with respect to the perturbation norm:

\beq\tag{44}
||\hatq||\equiv \{ \Delta I^*[\hatq]\}\tothehalf.
\eq

(41) implies, it may be noted, explicit bounds on the equilibrium potential vorticity gradient which signify conditions stronger than (31) that is required for {\it linear} stability. For {\it nonlinear} stability, the potential vorticity gradient not only needs to be positive (as for {\it linear} stability) but it must also be bounded from above.

\section{Discussion}
In this paper, a Hamiltonian formulation has been used to develop general theoretical results for zonal shear flows with the inclusion of the vortex stretching effect of the deformed free surface. One such result is a generalization of the FSW {\it zero angular momentum theorem} for localized nonlinear structures (on or off the $\beta$-plane). Off the $\beta$-plane, more general localized nonlinear structures with {\it non-zero} angular momentum are shown to be possible. On the other hand, even on the $\beta$-plane, localized nonlinear structures are shown not to be restricted by the FSW theorem provided they propagate at certain critical speeds. Another general theoretical result is concerned with {\it sufficient} conditions for {\it linear} and {\it nonlinear} stability in the Liapunov sense - the latter have been derived via explicit {\it convexity} conditions for the Casimir function and hence bounds on the equilibrium potential vorticity gradient. For {\it nonlinear} stability, the latter not only needs to be positive (as for {\it linear} stability) but it must also be bounded from above.

\section{Acknowledgments}
Our thanks are due to Dr. Leon Kamp for helpful discussions. This work has been supported by the J.M. Burgers Centre.

\end{document}